\documentstyle[preprint,tighten,aps,psfig]{revtex}

\begin{document}
\draft
\preprint{hep-ph/9706227, TTP97-20}
\title{Techniques for computing two-loop QCD corrections to $b \to c$
  transitions} 

\author{Andrzej Czarnecki and Kirill Melnikov}
\address{Institut f\"ur Theoretische Teilchenphysik, \\
Universit\"at Karlsruhe,
D-76128 Karlsruhe, Germany}
\maketitle

\begin{abstract}
We have recently presented the complete ${\cal O}(\alpha _s ^2)$
corrections to the semileptonic decay width of the $b$ quark at
maximal recoil.  Here we discuss various technical aspects of that
calculation and further applications of similar methods.  In
particular, we describe an expansion which facilitates the phase space
integrations and the treatment of the mixed real--virtual corrections,
for which Taylor expansion does not work and the so-called eikonal
expansion must be employed. Several terms of the
expansion are given for the ${\cal O}(\alpha _s ^2)$ QCD corrections
to the differential semileptonic decay width of the $b$--quark at
maximal recoil.  We also demonstrate how the light quark loop
corrections to the top quark decay rate can be obtained using the same
methods.  We briefly discuss the application of these techniques to
the calculation of the ${\cal O}(\alpha _s ^2)$ correction to zero
recoil sum rules for heavy flavor transitions.
\end{abstract}

\section{Introduction}

Precise determination of $|V_{cb}|$, 
a parameter of the Cabibbo-Kobayashi-Maskawa
(CKM) matrix,
is an important goal of many experimental studies. 
The current experimental limit \cite{RPP}
\begin{equation}
|V_{cb}| = 0.036 \mbox{  to  } 0.046 \quad (90\% \mbox{ CL})
\end{equation}
is based on measurements of the beauty hadron decays produced 
at the $\Upsilon(4S)$ resonance (by ARGUS
and by CLEO II) and in $Z$-boson decays  (by the four experiments at
LEP).  In the future large  samples of the $b$-hadrons collected at
$B$-factories (at SLAC and KEK) and at the hadron colliders will
increase the statistical accuracy to a few percent level.  To fully
exploit the anticipated experimental improvement, the theoretical
description of the $b$ decay must be known with comparable precision.

There are two methods of extracting the value of $|V_{cb}|$, based on
measurements of the exclusive decay $B\to \bar {D^\star} \bar l\nu_l$
and of the inclusive semileptonic decay width of $b$-hadrons $\Gamma
_{\rm sl}$.  These two methods rely on very different theoretical
considerations and experimental procedures and complement each other.
Their merits and theoretical uncertainties are summarized e.g.~in
Ref.~\cite{Shifman94,Neubert97}.  One of the major sources of the
theoretical error 
are the
perturbative QCD corrections at the two loop level.  For the exclusive
decays at the zero recoil point these corrections have recently been
calculated \cite{zerorecoil,zerorecoilA}.  This has significantly
improved the 
accuracy of the theoretical prediction for the exclusive method.

Regarding the inclusive method, recently the ${\cal O}(\alpha _s ^2)$
corrections to differential semileptonic decay width of the $b$--quark
at maximal recoil have been calculated \cite{MaxRec}.  Combined with
the previously obtained value at zero (minimal) recoil, these results
permitted an estimate of the complete ${\cal O}(\alpha _s ^2)$
correction to the total semileptonic decay width of the $b$--quark.

In the paper \cite {MaxRec} we  presented the results of
that calculation and discussed its phenomenological relevance.  The
purpose of the present paper is a detailed description of the methods
employed in that calculation.  We would like to note that a complete
calculation of two-loop corrections to a fermion decay width has never
been performed before, neither in QCD nor in QED (a longstanding
example are the two-loop QED corrections to the muon life
time). Therefore, in the calculation we describe here we had to go
beyond the traditional methods used in higher--orders calculation.
We hope that a description of some technical aspects and
methods will be of interest for the community.

Let us first mention the difficulties one encounters when trying to
compute the fourth order corrections to the fermion decays. 
One of the problems is an appropriate treatment of the real radiation of 
one or two gluons. 
 For  the virtual
radiative corrections there exists a number of algorithms, 
permitting an efficient, analytical treatment of 
a large number of complicated  
diagrams, which typically
appear in such calculations.
On the contrary, no similar algorithms were available so far for the 
treatment of the real radiation.

The reason why the real radiation at order ${\cal O}(\alpha^2_s)$ is
difficult to evaluate is that the particle in the initial state 
(the decaying $b$ quark) carries a color charge and therefore can radiate.
It is the presence of 
the massive propagator of the initial quark 
which makes the integrations over
the phase space very tough. 
The kinematical configuration, where the invariant mass of the
leptons is equal to zero and the quark
in the final state is massive is the first
case where a complete analytical evaluation of the real radiation
of two gluons in a decay of a fermion turns out possible.

Another potential 
problem is the treatment of  diagrams 
which represent  one--loop virtual corrections to a single
gluon emission in the $b$--quark decay.
The virtual corrections in such
situation are  one--loop.
Therefore one might naively expect that
this case does not require any sophisticated investigation. 
Unfortunately, the integration of the one--loop formulas (especially
of the boxes) over the three body phase space is difficult.
However, it turns out  possible to express the result for the loop 
as an
expansion which can be easily integrated over the phase space.  With
a systematic algorithm for the expansion, this approach shifts the
burden of the calculation to the computer.

The idea which permitted us to calculate the contribution of  the
real radiation of one or two gluons is (qualitatively speaking) 
the expansion in
the velocity of the final quark. 
In the limit $m_c \to m_b$ the charm quark in the final state is a
slowly moving particle, with spatial components of its 
momentum of the order of 
$m_b-m_c$, much smaller than its mass.
The momenta of gluons and of leptons (in the case when the invariant
mass of the leptons is zero) 
are also of the order of $m_b - m_c$. It turns out that by a proper
choice of the phase space variables  one can systematically expand the  
amplitudes and the phase space in 
$\delta \equiv (m_b-m_c)/m_b \ll 1$ (sometimes we also use an
equivalent expansion parameter $\beta\equiv
(m_b^2-m_c^2)/m_b^2=\delta(2-\delta)$; throughout this paper we use
$m_b$ as the unit of mass, putting $m_b=1$). The details of the phase
space parameterization will  be explained
in detail.

A word of caution is in order  here. Our technique proved to be very
useful in problems where the mass of the quark in the final state
does not differ too much from the mass of the quark in the initial
state. In this respect an ideal application are 
semileptonic $b \to c$ transitions.  On the other hand,
for such problems as muon  or top quark decay, the expansion
parameter may be close to unity and it is not clear  at present
if our procedure is of any use there.  We will,
however, show an example where our procedure 
remains meaningful and delivers reliable 
predictions even in the case when the expansion parameter equals unity.

The paper is organized as follows. In the next section we discuss how
the real radiation of two gluons can be computed. Section
\ref{sec:tensors} is devoted to the treatment of tensor
integrals. Then we present a detailed study of one--loop virtual
corrections to a single gluon emission in the $b$--decay. We show that
in such diagrams a new type of Feynman integrals appears and discuss
their evaluation. Section \ref{sec:applic} is devoted to the
applications of our techniques.  We present the result for the ${\cal
O}(\alpha _s ^2)$ correction to the differential semileptonic decay
width $b \to c\; l\; \nu _l$ up to $\delta ^{11}$. We also demonstrate
how the known light quark corrections to the top quark decay width can
be obtained with good accuracy if the expansion up to high powers of
$\delta $ is performed, and discuss an application of our technique to
corrections to the zero recoil sum rules.  In the last section we
present conclusions.

\section{Emission of two gluons in semileptonic {${b}$} decays}

We first consider the radiation of two real gluons in a semileptonic
$b$--decay 
$$b \to c + e + \nu _e + g_1+ g_2$$ 
and discuss how its contribution to the width can be computed.

Throughout this paper we denote the particles and their four-momenta
by the same letters, i.e.~the momentum of the $b$--quark is $b$
etc. Moreover, we denote the momentum of the lepton pair $e + \nu _e$
as $W$ and often speak about the decay $ b \to c +W$. All formulas in
this paper apply to the case when the invariant mass of the leptons is
zero, i.e. $W^2 = 0$, if not  stated otherwise.

Our aim is to construct an expansion in $\delta=(m_b-m_c)/m_b \ll 1$.
In this case the spatial momentum of the $c$ quark is of the order of
$m_b - m_c$.  The momenta of the lepton pair and of the gluons are
also of the order of $m_b - m_c$ because $W^2=g_1^2=g_2^2 =0$. The
smallness of these momenta permits an expansion in terms of $\delta$.

\subsection {Expansion of the propagators}
\begin{figure}[h]
\hspace*{-5mm}
\begin{minipage}{16.cm}
\[
\mbox{
\hspace*{10mm}
\begin{tabular}{cc}
\hspace*{10mm}
\psfig{figure=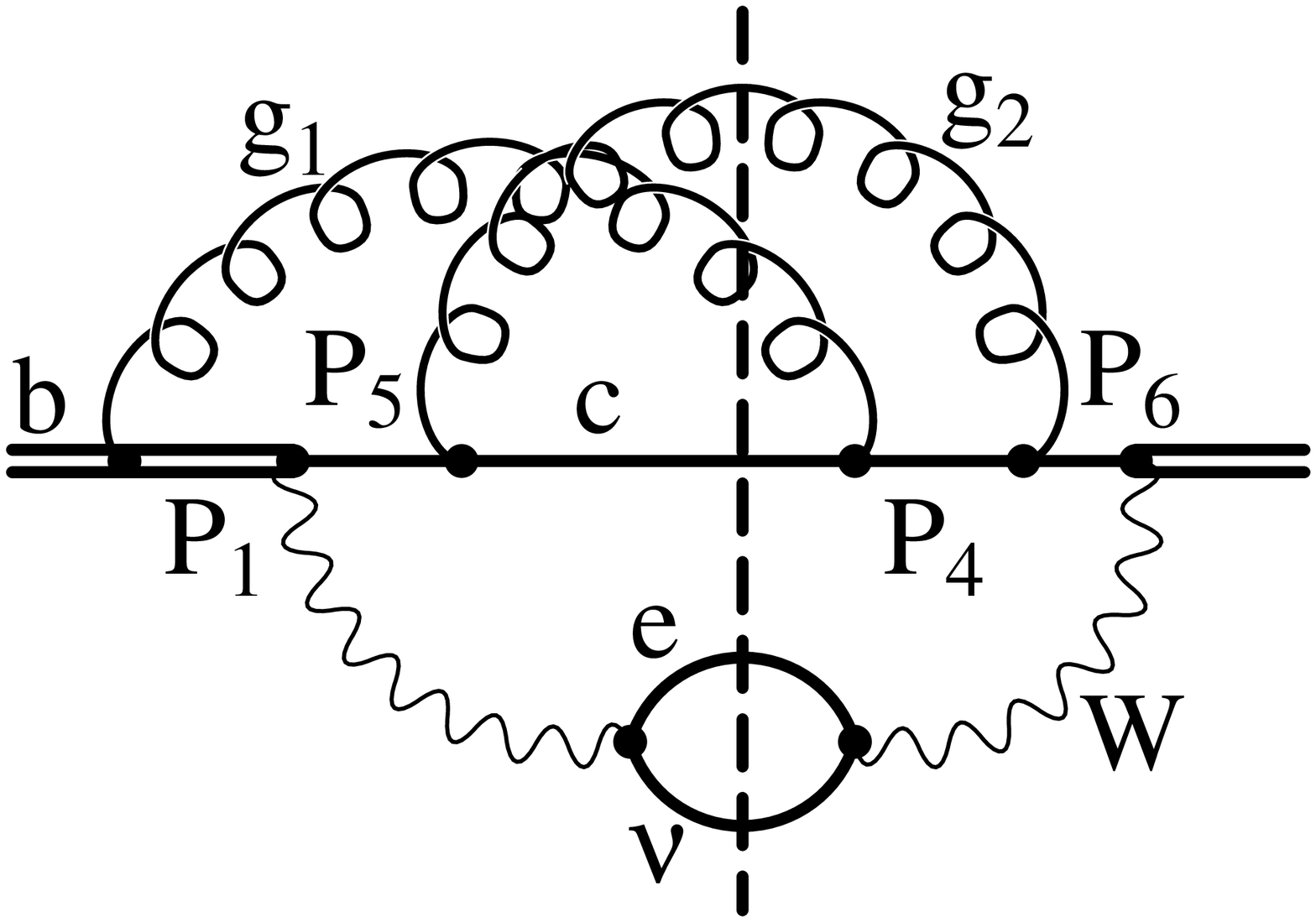,width=48mm,bbllx=210pt,bblly=410pt,%
bburx=630pt,bbury=550pt} 
&\hspace*{25mm}
\psfig{figure=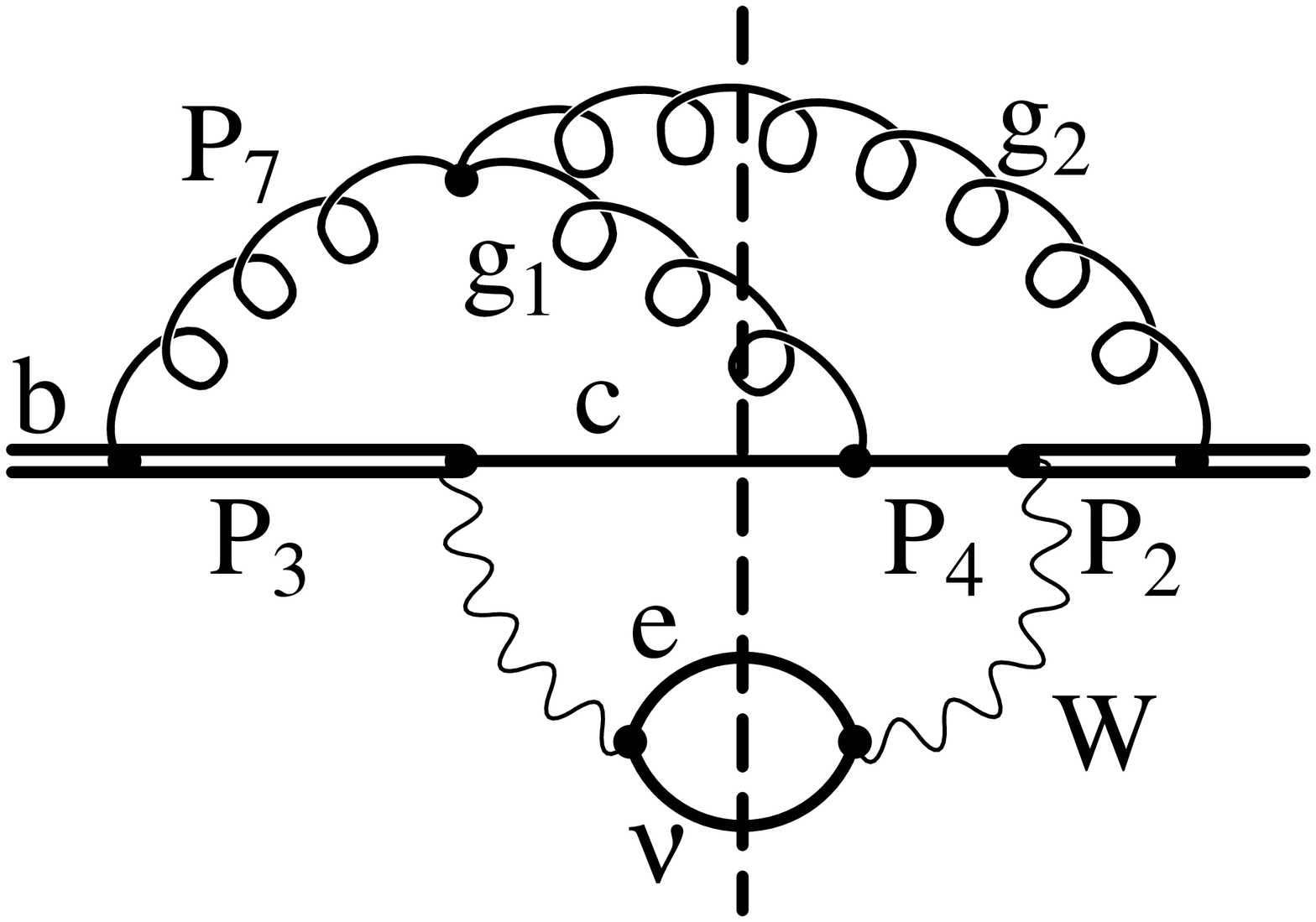,width=48mm,bbllx=210pt,bblly=410pt,%
bburx=630pt,bbury=550pt}
\\[25mm]
\hspace*{-12mm}
(a) & (b)
\end{tabular}}
\]
\end{minipage}
\caption{Notation used for the
  propagators $P_i$ in diagrams contributing to
  semileptonic $b$ decays: examples with two gluons (a) QED-like  and
  (b) with a nonabelian coupling. } 
\label{fig:propagators}
\end{figure}

To show that such an expansion is indeed possible,
we list here all 
propagators of the virtual particles which appear 
in a calculation of the two gluon emission in the semileptonic
$b$--decay:
\begin{eqnarray}
P_1 &=& \frac {1}{(b-g_1)^2-m_b^2}=\frac {1}{-2bg_1},
\qquad
P_2= \frac {1}{(b-g_2)^2-m_b^2}=\frac {1}{-2bg_2},
\nonumber \\
P_3&=& \frac {1}{(b-g_1-g_2)^2-m_b^2}=\frac {1}{-2b(g_1+g_2)+2g_1g_2}
\nonumber \\
P_4 &=& \frac {1}{(c+g_1)^2-m_c^2}=\frac {1}{2cg_1},
\qquad
P_5= \frac {1}{(c+g_2)^2-m_c^2}=\frac {1}{2cg_2},
\nonumber \\
P_6 &=& \frac {1}{(c+g_1+g_2)^2-m_c^2}=\frac {1}{2c(g_1+g_2)+2g_1g_2},
\qquad
P_7 = \frac{1}{(g_1+g_2)^2}=\frac {1}{2g_1g_2}
\end{eqnarray}
All these propagators are shown in the examples in
Fig.~\ref{fig:propagators}. 
If we eliminate the momentum of the $c$ quark using momentum
conservation,  $c=b-W-g_1-g_2$, we can expand the propagators $P_4...P_6$
in the Taylor series with respect to the ``small'' momenta $g_1$,
$g_2$, $W$. 
Explicitly, we have
\begin{eqnarray}
P_4 &=& \frac {1}{2bg_1}\sum _{j=0}^{\infty} 
\frac {(2g_1W+2g_1g_2)^j}{(2bg_1)^j},    \qquad
P_5 = \frac {1}{2bg_2}\sum  _{j=0}^{\infty} 
\frac {(2g_2W+2g_1g_2)^j}{(2bg_2)^j},
\nonumber \\ 
P_6 &=& \frac {1}{2b(g_1+g_2)-2g_1g_2}
\sum  _{j=0}^{\infty} 
\frac {[2(g_1+g_2)W]^j}{[2b(g_1+g_2)-2g_1g_2]^j}.
\end{eqnarray}
From these expressions we see that the expansion does not generate
any new infrared divergences and, therefore, appears to be permissible.
After this expansion only four
different types of denominators remain. We list them for clarity:
\begin{eqnarray}
P_1^0 &\equiv&\frac {1}{2bg_1}, \qquad
P_2^0\equiv \frac {1}{2bg_2},
\nonumber \\
P_3^0 &\equiv&\frac {1}{2b(g_1+g_2)-2g_1g_2},\qquad
P_7 = \frac {1}{2g_1g_2}.
\end{eqnarray}
As will become clear from the discussion below, 
it is more convenient to
use $P_3^0$ as one of the basic propagators than to expand it any further.

Let us emphasize at this point the advantages of the above
expansion. As can be seen from the above expressions, the 
four--momentum of lepton pair $W$ and the momentum of the $c$--quark do
not appear in the denominator. This immediately implies that
the integral over the phase space of $W$ and $c$ factorizes. 
Therefore, the expansion suggests a simple way how the rather
nontrivial integration over the phase space of 
four particles in the final state can be
reduced to a more familiar case of the three body phase space.
This reduction is described in the next section.

\subsection{Phase--space integration for the emission of two gluons}
\label{sec:pstwo}
After we have checked that the propagators of all virtual
particles can be expanded, the same must be
demonstrated for the phase space element. This is done in the
present section.

Considering the semileptonic decay of the $b$--quark and integrating first
over the lepton phase space, one obtains the following phase space
integration element
(we consider only the point where the invariant mass of the lepton pair
$W^2$ is zero):  
$$
{\rm d} R_4 =  [dc][dW][dg_1][dg_2] \delta^D (b-c-g_1-g_2-W),
$$
where a shorthand notation is
$$
[dp] = \frac {{\rm d}^{D-1}p}{2p^0},\qquad D=4-2\epsilon.
$$
It is convenient to introduce 
an auxiliary vector $H$ which equals to the sum of the momenta
of the $c$--quark and $W$ boson, 
$H \equiv c+W$. 
With this notation the phase space integral becomes
$$
dR_4 = \int  dH^2 \int [dH] [dg_1][dg_2] 
\delta^D (b-H-g_1-g_2) \int [dW][dc] \delta^D (H - W -c).
$$
The integration over the $(W,c)$ phase space gives (for now we assume
the integrand contains no  $W$ in the numerator; the tensor integrals
with $W$ dependence will be analyzed in section \ref{sec:tensors}) 
$$
R_2^{Wc} =\int [dW][dc] \delta^D (H - W -c)= 
\frac {\Omega _{D-1}}{2^{D-1}} \frac {(H^2-m_c^2)^{D-3}}{(H^2)^{D/2-1}}
$$
where
$$
\Omega _{D-1} = \frac {2~~\pi ^{3/2-\epsilon}}{\Gamma (3/2 - \epsilon)}
$$
is the volume of a $D-1$ dimensional sphere of a unit radius.

Having performed the integration over the $(W,c)$ phase space we are left 
with the three particle phase space of one massive ($H$) and two massless
($g_1$, $g_2$)
particles and the integration over the square of the momentum $H$. 

We use the variables (with $m_b^2 \equiv 1$)
\begin{equation}
H^2 = y,\qquad (g_1+g_2)^2 = u,\qquad(H+g_2)^2 = z,
\end{equation}
and get the following expression for the four particle phase space
\cite{topfermion}
\begin{equation}
{\rm d} R_4 = \Bigg ( \frac {\Omega _{D-1}}{2^{D-1}} \Bigg )^3 
\frac {1}{B(1-\epsilon,1-\epsilon) }
\int {\rm d} y {\rm d} z {\rm d} u u^{-\epsilon}z^{-\epsilon}(u_m-u)^{-\epsilon}
\frac {(y-m_c^2)^{1-2\epsilon}}{y^{1-\epsilon}},
\end{equation}
with $u_m=(1-z)(z-y)/z$ and the following limits of integrations
\begin{equation}
m_c^2 < y < 1,\qquad y < z <1,\qquad 0 < u < u_m.
\end{equation}
In order  to expand the phase space element in powers of
$\beta=(m_b^2-m_c^2)/m_b^2$ it is useful to  change the
variables in the above expression
\begin{eqnarray}
y &=& 1-\beta x_1,\qquad z=1-\beta
x_1x_2,\qquad u=u_mx_3, 
\nonumber \\
u_m&=&\frac {\beta^2x_1^2x_2(1-x_2)}{1-\beta x_1x_2}.
\end{eqnarray}
The integration limits for all $x_i$ are the same:
$$
0 < x_i < 1
$$
Expressing the phase--space element through these variables and
restoring proper dimensionality we find:
$$
{\rm d} R_4 = R_4^0~\beta ^{5-6\epsilon} \int  _{0}^{1}
{\rm d} x_1 {\rm d} x_2{\rm d} x_3 J(x_1,x_2,x_3)
\Bigg [\frac {1}{(1-\beta x_1 x_2)^{1-\epsilon}} 
\frac {1}{(1-\beta x_1)^{1-\epsilon}} \Bigg ],
$$
where
\begin{eqnarray}
J(x_1,x_2,x_3) &=&
x_1^{3-4\epsilon}(1-x_1)^{1-2\epsilon}
x_2^{1-2\epsilon}(1-x_2)^{1-2\epsilon}x_3^{-\epsilon}(1-x_3)^{-\epsilon},
\nonumber \\
R_4^0 &=& \frac {(m_b^2)^{2-3\epsilon}}{B(1-\epsilon,1-\epsilon)}
 \Bigg  ( \frac {\Omega _{D-1}}{2^{D-1}} \Bigg )^3.
\end{eqnarray}

Our final aim is to expand ${\rm d} R_4$ in the limit of $m_c\to m_b$.
The above expression
is well suited for this purpose.
Expanding it in powers of $\beta = (m_b^2-m_c^2)/m_b^2$
gives rise to simple integrals which can be expressed by Beta
functions: 
\begin{eqnarray}
\lefteqn{ \int {\rm d} x_1~{\rm d} x_2~{\rm d} x_3 J(x_1,x_2,x_3) x_1^n x_2^m x_3^k }
\nonumber \\ &&\qquad\qquad=
B(n+4-4\epsilon,2-2\epsilon) B(m+2-2\epsilon,2-2\epsilon) B(k+1-\epsilon,1-\epsilon).
\label{eq:master}
\end{eqnarray}

\subsection {Basic propagators}

The basic propagators 
$P_1^0$, $P_2^0$, $P_3^0$, and $P_7$, 
expressed in terms of the variables $\{x_i\}$
yield simple expressions which can be integrated over the phase space.
We find
\begin{eqnarray}
P_1^0 &=& \frac {1}{2bg_1} = \frac {1}{\beta x_1 x_2}, \qquad
P_2^0 = \frac {1}{2bg_2} = \frac {1-\beta x_1 x_2}
{\beta x_1(1-x_2) [1-\beta x_1x_2(1-x_3)]},
\nonumber \\ 
P_3^0 &=& \frac {1}{2b(g_1+g_2)-2g_1g_2} = \frac {1}{\beta x_1},\qquad 
P_7 = \frac {1}{2g_1g_2} = 
\frac {1-\beta x_1 x_2}{\beta^2 x_1^2 x_2 (1-x_2) x_3}.
\end{eqnarray}
Clearly, the expansion of these quantities in terms of small $\beta$
is readily performed. The resulting integrals are very
similar to the integrals in eq.~(\ref{eq:master}) and can be expressed in 
terms of the Beta functions.

The same is true for all scalar products of the four
momenta which enter the calculation. Therefore, the above discussion
demonstrates that the part of the semileptonic decay width of the
$b$--quark containing radiation of two gluons can be 
calculated by expanding the matrix element and the four particle phase
space in powers of $\beta$.
What is still missing in our discussion is the study of the
integrals over the $(W,c)$ phase space when the numerator depends on
$W$. We discuss this issue in the next section.

\section{Tensor integrals}
\label{sec:tensors}
The expansion of the propagators in terms
of the momenta $W$, $g_1$, $g_2$, may result in
high powers of the scalar products 
$(Wg_1)^{a}(Wg_2)^{b}$, which  have to be integrated over the $(W,c)$ 
phase space. In this section we present efficient
methods for such integrations.  
The results
of this section are not restricted to the case $W^2 = 0$ and 
are likely to be useful for solving other problems.

The main object we are going to discuss is the following integral
over the $(W,c)$ phase space:
\begin{eqnarray}
I(a,b) &=&  \int [dW][dc] \delta^D (H - W -c) (Wg_1)^a (Wg_2)^b 
\nonumber \\
& \equiv &
\langle  (Wg_1)^a (Wg_2)^b \rangle _{H=W+c}
\end{eqnarray}
for arbitrary $a,b > 0$. 
This integral can be rewritten as
\begin{eqnarray}
I(a,b) &=& g_1^{\mu_1}....g_1^{\mu _a}~ 
           g_2^{\mu_{a+1}}....g_2^{\mu_{a+b}}~
T_{\mu,a+b}
\nonumber \\
T_{\mu,n} &\equiv &
\langle W_{\mu_1}....W_{\mu _n} \rangle _{H=W+c}.
\end{eqnarray}
The important property of the tensor $T$ is that it depends on 
a single external vector only ($H$).  This property simplifies the
calculation of the necessary integrals.  We discuss
below two approaches to such calculation; their relative merits
depend on the capabilities of the symbolic manipulation languages, if
the algorithm is to be implemented using a computer.

\subsection {Method 1}

The first approach is more general and potentially more
efficient.  To fully exploit the dependence on the single
external momentum we
can extract the component of the vector 
$W$ which is parallel to it. For this purpose we write:
\begin{equation}
W = W_\bot + c_1H,\qquad W_\bot H=0,\qquad ~c_1=\frac {WH}{H^2} 
\end{equation}
The value of $c_1$ is then fixed, because in the general case $W^2 \ne
0$ we have 
\begin{equation}
WH = \frac {1}{2}(H^2+W^2-m_c^2).
\end{equation}
Therefore, if we use the above substitution for $W$ in  
the tensor integral $T^{\mu,n}$, we obtain a sum of
tensor integrals over transverse components of
the $W$ (with respect to $H$):
\begin{equation}
T_\bot ^{\mu,n}\equiv \langle 
W_\bot^{\mu_1}.....W_\bot^{\mu_n} \rangle _{H=W+c}.
\end{equation}
Evidently, the tensor structure of this integral is trivial:
since $W_{\bot}H = 0$, 
$T_\bot ^{\mu,n}$
must be proportional to the
{\it absolutely symmetric tensor} of the $(D-1)$ dimensional space.
The ``building block'' is the metric tensor of $(D-1)$ dimensional space:
$g_{\mu\nu}-H_{\mu}H_{\nu}/H^2$. 
We note that only tensors of  {\it even} 
rank contribute. 
The generation of the absolutely symmetric tensor
of $D-1$ dimensional space can be easily encoded in a 
symbolic manipulation program.  Hence, the algorithm described above
permits an efficient 
treatment of the tensor integrals which appear in the problem at hand.

\subsection{Method 2}
Using the same idea as in the previous section we can advance
slightly further with the analytical calculation of the integral $I(a,b)$.

After writing $W = W_\bot +c_1H$ and noticing that $g_1H$ and $g_2H$
are constants in the $(W,c)$ phase space, we conclude that 
the non-trivial integrals to be computed are
\begin{equation}
I_{\bot}(a,b)\equiv \langle (W_\bot g_1)^a (W_\bot g_2)^b \rangle
_{H=W+c} =  
\langle (W_\bot g_{1\bot})^a(W_\bot g_{2\bot})^b \rangle _{H=W+c},
\end{equation}
with $g_{i\bot}$ (i=1,2)  defined in analogy to $W_\bot$.

The integral $I_{\bot}$ is a Lorentz--scalar and we can choose an
arbitrary frame for its calculation.
For instance, we consider $H$ to have only the zeroth component and
put the $z$--axis of the $D-1$ dimensional space along the vector
$g_{1\bot}$. Then, rewriting $g_{2\bot}$ as 
\begin{equation}
g_{2\bot}=g_{2a}+c_2g_{1\bot},\qquad g_{2a}g_{1\bot}=0,
\qquad c_2=g_{2\bot}g_{1\bot}/g_{1\bot}^2,
\end{equation}
we find that $I_\bot$ is expressed as a sum of integrals of the from:
\begin{equation}
I_1(a,b) = 
\langle (W_\bot g_{1\bot})^a (W_\bot g_{2a})^b \rangle _{H=W+c}.
\end{equation}
This integral is non-zero only  if both  $a$ and $b$ 
are even integers.

Since $g_{2a}g_{1\bot}=0$, we can put the $x$-axis of the $D-1$
dimensional space along the vector $g_{2a}$. Then we have:
\begin{eqnarray}
I_1(a,b)&=&R_2^{Wc} |W_\bot|^{(a+b)}|g_{1\bot}|^a|g_{2a}|^b I_{2}(a,b)
\nonumber \\
I_2(a,b)&=&
\int \frac {d\Omega _{D-1}}{\Omega _{D-1}} 
\Bigg ( \frac {W_\bot g_{1\bot}}{|W_\bot||g_{1\bot}|} \Bigg )^a
\Bigg ( \frac {W_\bot g_{2a}}{|W_\bot||g_{2a}|} \Bigg )^b.
\end{eqnarray}
The integration element is
\begin{equation}
{\rm d}\Omega _{D-1}= \prod_{i=1}^{D-2}\sin^{D-2-i}\theta_i ~{\rm d}\theta _i,
\qquad 0 < \theta _i < \pi \quad (i\ne D-2), 
\qquad 0 < \theta _{D-2} < 2\pi. 
\end{equation}
If we choose $z$ and $x$ axes as described above, we get
\begin{eqnarray}
\frac {W_\bot g_{1\bot}}{|W_\bot||g_{1\bot}|} &=& - \cos \theta_1,
\nonumber \\
\frac {W_\bot g_{2a}}{|W_\bot||g_{2a}|} &=& - \sin \theta _1 
\cos \theta _2. 
\end{eqnarray}
The minus sign in the above equations arises because
the transverse vectors here are spacelike.
In calculating $I_2(a,b)$ the following formula is useful:
\begin{equation}
\int  _{0}^{\pi} {\rm d}\theta \sin ^\alpha\theta  \cos^\beta \theta 
=B\left( \frac {1+\alpha}{2},\frac {1+\beta}{2} \right).
\end{equation}
This formula is valid for  even  and positive $\beta$. 
Using it we get the result for the integral 
$I_2(a,b)$:
\begin{equation}
I_2(a,b)= { (-1)^{a+b} 
\Gamma\left( {3\over 2}-\epsilon\right)
\Gamma\left( {1+a\over 2}\right)
\Gamma\left( {1+b\over 2}\right)
\over
\pi \Gamma\left( {a+b+3-2\epsilon \over 2}\right)
}
\end{equation}
In order to use these results for the further integration
over the phase space $(H,g_1,g_2)$
it is necessary to express the modulus of the vectors 
$W_\bot$, $g_{1\bot}$, $g_{2a}$ 
in terms of the variables $\{x_i \}$ introduced in the previous
section. This  is a cumbersome but  straightforward task and we do not
discuss it here in any detail.

The steps described above allow an easy
implementation in any symbolic manipulation program. They
provide an efficient and uniform treatment of the tensor integrals
which appear in this calculation.

\section{One loop virtual correction and  one real gluon}

Another source of the $O(\alpha _s ^2)$ corrections are the one--loop
virtual corrections accompanied by a radiation of an additional real
gluon.  Again, we would like to expand those diagrams in powers of
$\delta = (m_b-m_c)/m_b$.  The purpose of this section is to provide
a suitable algorithm.

There is
a principal difference from the radiation of two real gluons
in this part of the corrections. 
Namely, it is not sufficient here to perform a Taylor
expansion of the relevant diagrams. 

\begin{figure}[h]
\hspace*{-5mm}
\begin{minipage}{16.cm}
\[
\mbox{
\hspace*{10mm}
\psfig{figure=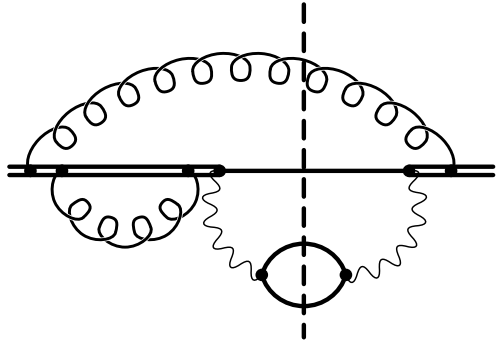,width=50mm,bbllx=210pt,bblly=410pt,%
bburx=630pt,bbury=550pt} 
}
\]
\end{minipage}
\vspace*{23mm}
\caption{The simplest diagram in which the Taylor expansion in
  $(m_b-m_c)/m_c$ leads to spurious divergences and 
 must be supplemented by an eikonal expansion.} 
\label{fig:simplest}
\end{figure}

To see this, let us analyze
a simple example. 
We consider a diagram where the $b$--quark radiates a gluon
and then a one--loop self--energy correction is inserted
on the $b$ quark line (see Fig.~\ref{fig:simplest}).
Taking the limit
$m_b \to m_c$ in such
situation 
implies that
the momentum of the real gluon goes to zero. Therefore the
self--energy diagram should be calculated close to the mass shell
of the $b$--quark. 
It
is well--known
that in this limit the self--energy diagram has an on-shell
logarithmic singularity
of the form 
$$
\log(M^2-p^2) \sim \log(M^2-m^2) \sim \log(\delta).
$$

Since this diagram contains $\log(\delta)$ it is clear that it cannot
be calculated as a Taylor expansion with respect to $\delta$.
Therefore, a more sophisticated procedure is needed and the method of
{\it eikonal expansions} \cite {Smirnov96,CzarSmir96} is used in this
case.  We illustrate this approach in the following example.

\subsection {An example of the eikonal expansion}

In this subsection we  consider a simple example 
of a one--loop integral, where a result obtained from the eikonal expansion
can be compared with an exact formula.  We calculate the scalar
self--energy diagram:
\begin{equation}
B(p^2,m^2) = \int \frac {{\rm d}^Dk}{(2\pi)^D} \frac {1}{k^2[(k+p)^2-m^2]}.
\label {example}
\end{equation}
Performing Feynman parameterization, integrating over the loop momenta,
and expanding in $\rho = (m^2-p^2)/p^2$
we get (using $p^2 =1$ here):
\begin{equation}
B(p^2,m^2) = \frac {i\Gamma (1+\epsilon)}{(4\pi)^{D/2}}
\left( \frac {1}{\epsilon}+2 -\rho + \rho \ln \rho -\frac {\rho ^2}{2}
+\frac {\rho ^3}{6} -\frac {\rho ^4}{12}+\ldots\right)
\end{equation}

The appearance of the $\log (\rho)$ term in this result
signals that the Taylor expansion  of the {\em integrand} will be
insufficient. 
In order to get the correct result  we add
the eikonal expansion.  For this purpose we write
\begin{equation}
B(p^2,m^2) = B_{\rm t} + B_{\rm eik}
\end{equation}
The Taylor expansion term $B_{\rm t}$ is obtained by expanding the
integrand  
in (\ref {example}) in a Taylor series in $\rho$. This yields a set
of one--loop on--shell integrals which are well known.
The validity of such an expansion 
is determined by the condition $k^2+2pk \gg \rho p^2$. If this is not the
case, the Taylor expansion breaks down.  
This breakdown gives rise to an
infrared singularity 
which occurs in the region
where $\rho p^2\sim 2pk$ and $k^2\ll \rho p^2$. 
This 
spurious divergence can be cancelled by adding an expansion of the
integrand in $k^2$ (for a more detailed discussion see
\cite{Smirnov96,CzarSmir96}). 

One should note in addition that if any power of $k^2$ appears
in the numerator we get an integral of the form:
$$
 I(n) = 
\int \frac {{\rm d}^Dk}{(2\pi)^D} \frac {(k^2)^n}{2pk-p^2\rho},
\qquad ~~~ n \ge 0.
$$
An important point is that such integrals are
zero within the dimensional regularization  framework.  One simple reason
for this is
that the integral over transverse components of $k$ is scaleless.

Hence, the only term in the eikonal expansion we should 
consider is the one with $k^2$  neglected in the second 
propagator. The eikonal integral therefore reads:
\begin{equation}
B_{\rm eik} = \int \frac {{\rm d}^Dk}{(2\pi)^D} 
\frac {1}{k^2(2pk -p^2 \rho)}.
\end{equation}
It is evident that this integral scales with $\rho$ as 
$B_{\rm eik} \propto \rho ^{1-2\epsilon}$. It is this dependence on $\epsilon$
which  gives $\log(\rho)$ in the final result.

Consider $B_{\rm eik}$ in the rest frame of $p$. Performing 
the integration over the time component of the loop momenta first, one gets
a simple integral representation over the transverse components of the
loop momenta which can be easily calculated. More sophisticated eikonal
integrals are considered in the next section, where all necessary
formulas can be found.

\subsection{Eikonal integrals: preliminaries}
\label{eikogen}

Here we provide a
set of formulas for dealing with integrals appearing in 
the eikonal expansions.  As is clear from the example presented
above, the region of integration in which we are interested  here
is characterized by the  values of the virtual momentum of the
order of the momentum of the real gluon emitted in the decay.

We denote the loop momentum by $k$ and the momentum of the
real gluon by $g$; explicitly, the eikonal expansion of the propagator
$$
P(k) = \frac {1}{(k+b-g)^2-m_b^2}=\frac {1}{k^2+2k(b-g)-2bg}
$$
is given by
$$
P(k)_{\rm eik}=\frac {1}{2kb-2bg}\sum_{0}^{\infty} \left( 
\frac {2kg-k^2}{2kb-2bg} \right) ^{j}.
$$

The basic  eikonal integral in the one--loop correction then 
reads\footnote{In the Feynman diagrams with self--gluon couplings the
situation is more complicated. Their treatment is described in
subsection \ref{sgc}.}
\begin{equation}
I^{\pm}(n,m,a,b)=\int \frac {{\rm d}^Dk}{(2\pi)^D}
\frac {(kW)^a(kg)^b}
{(k^2+i\delta)(2kQ \pm 2Qg +i\delta)^n (2kQ + i\delta)^m}
\label{eik}
\end{equation}

In the actual calculation $Q$ is always a {\it time--like} vector which
for different diagrams can be $b,~b-g,~c+g$.

As we pointed out above, the integrals with  
any power of $k^2$  in
the numerator vanish.  There are several ways to see this. First, the
integrals over transverse (with respect to $Q$)
components of the loop momenta are scaleless; also
if there is no $k^2$ in the denominator,
the poles of
the integrand are located only on one side of the
$k^0$ integration axis. Also, if $n < 0$, the integral 
in eq.~(\ref{eik}) is zero because it is scaleless.

Let us describe the most economic way to 
calculate $I^{\pm}$. 
We choose the 
Lorentz frame in which $Q=(Q^0,0,\ldots,0)$.
The integration over the time-like component of the vector $k$ can 
be performed using the residues.  If the integration
contour is closed in the upper half--plane, only that pole in $1/k^2$
contributes for which  $k_0 = -k_\bot$.

After $k_0$, the integration over the $D-1$ dimensional space should be
performed. 
To discuss the  integration over $k_\bot$ 
we first note that after integrating
over $k_0$ the denominator of the integral $I^{\pm}(n,m,a,b)$ depends
on $k_\bot ^2$ only. Therefore, all tensor integrals
which appear due to the numerator structure are simply related to
the $D-1$ absolutely symmetric
tensor of the corresponding rank.
Hence, the tensor integrals here are similar to the ones discussed in
Section \ref{sec:tensors}.

Finally, we are left with the radial 
integration over the modulus of $k_{\bot}$.
The following formulas are useful:
\begin{eqnarray}
I^{-}(n,m,m_1)&=&\int \frac {{\rm d}^Dk}{(2\pi)^D}
\frac {(k_\bot^2)^{m_1/2}}
{(k^2+i\delta)(2kQ - 2Qg +i\delta)^n (2kQ + i\delta)^m}
\nonumber \\
&=&\frac {i\Omega _{D-1}}{(2\pi)^{D-1}}B(2-m+m_1-2\epsilon,n+m-m_1-2+2\epsilon)
\nonumber \\
&&
\times  (Qg)^{2-m+m_1-n-2\epsilon}(Q^2)^{-1-m_1/2+\epsilon}
            (-1)^{n+m+m_1+1}2^{-m-n-1}
\nonumber \\
I^{+}(n,m,m_1) &=& \int \frac {{\rm d}^Dk}{(2\pi)^D}
\frac {(k_\bot^2)^{m_1/2}}
{(k^2+i\delta)(2kQ + 2Qg +i\delta)^n (2kQ + i\delta)^m}
\nonumber \\
&=&\frac {i\Omega _{D-1}}{(2\pi)^{D-1}}
(Qg)^{2-m+m_1-n-2\epsilon}(Q^2)^{-1-m_1/2+\epsilon}
\nonumber \\
&&\times \frac {(-1)^{n+m+m_1+1} 2^{-m-n-1}}  {(n-1)!}
~\frac {\Gamma(2-m+m_1-2\epsilon)}{\Gamma(3-n-m+m_1-2\epsilon)}
\nonumber \\
&&\times \Big [\psi(2-m+m_1-2\epsilon)- \psi(-1+m-m_1+2\epsilon) \Big ],
\end{eqnarray}
with
\begin{equation}
\psi(z) = \frac {1}{\Gamma (z)}
\frac {{\rm d}\Gamma (z)}{{\rm d}z}.
\end{equation}
In the above formulas  $\Omega _{D-1}$ is the volume of the $D-1$
dimensional sphere of unit radius.

There is one subtle point concerning the expressions presented above.
Namely, it is easy to see that
the integrals $I^{+}$
have an additional singularity after the integration over $k_0$ has been 
performed. This singularity corresponds to the appearance of the
imaginary part in some of Feynman diagrams which contribute
to the result. Below we explain how this 
singularity was treated.

In the singular integrals 
the integration over the modulus of the transverse
momenta  is
$$
I^{+} \sim \int_{0}^{\infty} {\rm d} k_\bot 
\frac {k_\bot^{\alpha}}{(k_\bot -1-i\delta)^n}.
$$
The singularity is located at the point $k_\bot =1$. To make this
integral meaningful we rewrite it as
$$
I^{+} \sim \left. \frac {1}{(n-1)!} \left(\frac
{{\rm d}}{{\rm d} x}\right)^{(n-1)}   
\int _{0}^{\infty}  {\rm d} k_\bot
\frac {k_\bot^{\alpha}}{k_\bot -x-i\delta}\right|_{x=1}.
$$
Rescaling $k_\bot \to x~k_\bot $ we get:
$$
I^{+} \sim \left. \frac {1}{(n-1)!} 
\left(\frac {{\rm d}}{{\rm d} x}\right)^{(n-1)} x^{\alpha} 
\int_{0}^{\infty} {\rm d} k_\bot
\frac {k_\bot^\alpha}{k_\bot -1-i\delta}\right|_{x=1}.
$$
For the derivative we use
$$
\left(\frac {{\rm d}}{{\rm d} x}\right)^{(n-1)} x^{\alpha} = 
\frac {\Gamma (1+\alpha)}{\Gamma (2+\alpha -n)} x^{\alpha -n+1}.
$$
Finally, after performing the remaining integration over $k_\bot$ using
$$
{\rm Re} \int _{0}^{\infty} {\rm d} x \frac {x^\alpha}{x-1+i\delta}=
\psi(1+\alpha)-\psi(-\alpha),
$$
we arrive at the formula for $I^{(+)}$ quoted above.

\subsection{Eikonal integrals for the graphs with the gluon self--coupling}
\label {sgc}

The graphs with a triple gluon coupling lead to the most difficult
eikonal integrals. The difficulty originates from the fact that 
in such graphs one has two massless (gluon) propagators. 
A typical integral reads
$$
\int \frac {{\rm d}^Dk}{(2\pi)^D} 
\frac {1}{k^2~(k-g)^2~((k+b-g)^2-m_b^2)((c+k)^2-m_c^2)}.
$$

The Taylor expansion of such integral is easy. However, in the eikonal
expansion one cannot expand massless propagators.
Therefore, the master integral in this case is
$$
I(n,m)=\int  \frac {{\rm d}^Dk}{(2\pi)^D} 
\frac {1}{k^2~(k-g)^2 (2kQ-2Qg)^n (2kQ)^m},
$$
where $Q$ is again a ``large'' time-like vector, as described in the 
previous section.

Note also, that in this case the terms with $k^2$ or $(k-g)^2$ in 
the numerator do contribute to the result.

Let us describe in detail how such integrals can be calculated.
For this purpose we first combine the massless denominators:
$$
\frac {1}{k^2~(k-g)^2} = \int _{0}^{1} \frac {{\rm d} x}{(k^2-2kgx)^2}=
 \int _{0}^{1} \frac {{\rm d} x}{((k-gx)^2)^2}
$$
Next,  we  decompose the product of two
eikonal propagators 
$$
\frac {1}{(2kQ-2Qg)^n(2Qk)^m}
$$
using partial fractions.
The integral $I(m,n)$  becomes now a sum of the integrals of
the following type:
\begin{eqnarray}
I_1(n)&=& \int _{0}^{1}~{\rm d} x~\int \frac {{\rm d}^Dk}{(2\pi)^D} 
\frac {1}{[(k-gx)^2]^2 (2kQ-2gQ)^n},
\nonumber \\
I_2(n)&=&\int _{0}^{1}~{\rm d} x~\int \frac {{\rm d}^Dk}{(2\pi)^D} 
\frac {1}{[(k-gx)^2]^2 (2kQ)^n}.
\end{eqnarray}

In both integrals we shift the integration momentum $k \to k + xg$ and
get 
\begin{eqnarray}
I_1(n)&=&\int _{0}^{1}~{\rm d} x~\int \frac {{\rm d}^Dk}{(2\pi)^D} 
\frac {1}{(k^2)^2 [2kQ-2gQ(1-x)]^n},
\nonumber \\
I_2(n)&=&\int _{0}^{1}~{\rm d} x~\int \frac {{\rm d}^Dk}{(2\pi)^D} 
\frac {1}{(k^2)^2 (2kQ+2gQx)^n}.
\end{eqnarray}
Now, in the first integral one makes the change of variables $x \to 1-x$
and then rescales the integration momenta $k \to xk$ in both $I_1$ and
$I_2$.  The integration over $x$ factorizes and can be done using
$$
\int _{0}^{1} {\rm d}x\; x^{\alpha}=\frac {1}{\alpha +1}.
$$

The integrals over $k$  differ from the integral in the
previous section by the presence of two 
powers of $k^2$ in the denominator.
In this case it is useful to perform a
transformation (similar to  
the technique of integration by parts \cite{che81}),
which reduces the integrals with the second power of $k^2$
in the denominator to the integrals with the first power only. 

To write down this transformation we introduce the notation
$$
I^{\pm}(\alpha,n,m)=\int \frac {{\rm d}^Dk}{(2\pi)^D} 
\frac {1}{(k^2)^\alpha (2kQ \pm 2Qg)^n (2kQ)^m}.
$$
Then
$$
I^{\pm}(2,n,m)=Q^2~\Big [
-2(m+1)I^{\pm}(1,n,m+2)-2~n~I^{\pm}(1,n+1,m+1) \Big ].
$$
Using this formula we end up with integrals with the first power of
$k^2$ in the denominator, for which
the formulas of the previous section are applicable.

\subsection{Integration over the single gluon phase space}

In the treatment of the radiation of one gluon with 
${\cal O}(\alpha _s ^2 )$ accuracy, the last step one has to do
is to perform the integration over the phase space of the real gluon.
The eikonal expansion, described above in detail, gives
the virtual corrections
in terms of powers and logarithms of the small parameters.
This form of the intermediate result 
simplifies the final integration over the 
phase space of the decay products.  This would not be the case if we
had an {\em exact} result for the loop: that would certainly contain
dilogarithms of complicated arguments.

The phase space integration with a single gluon is 
in its general structure
similar to the case of two real gluons, discussed in section
\ref{sec:pstwo}. 
For completeness, we give a short account of necessary formulas.

In the present case the phase space element is:
\begin{eqnarray}
{\rm d} R_3 &=& \int [dc][dW][dg]\delta^D (b-c-g-W),
\nonumber \\
&=& \left(\frac {\Omega _{D-1}}{2^{D-1}} \right)^2 m_b^{2-4\epsilon}
\beta ^{3-4\epsilon}
\int _{0}^{1} {\rm d} x J(x) \frac {1}{(1-\beta x)^{1-\epsilon}}
\nonumber \\
J(x) &=& x^{1-2\epsilon}(1-x)^{1-2\epsilon}
\end{eqnarray}
with
\begin{eqnarray}
H&=& W+c,\qquad H^2 = 1-\beta x,
\nonumber \\
bH&=& 1-\frac {\beta x}{2},\qquad bg ~=~Hg~=\frac {\beta x}{2}.
\end{eqnarray}

The master integral in this case reads:
\begin{equation}
\int_{0}^{1} {\rm d} x J(x) x^n =  B(n+2-2\epsilon,2-2\epsilon).
\end{equation}
The results of the eikonal integrals require a slight modification of
the master integral.
As we saw in  section \ref{eikogen}, those results contain 
$(bg)^{-2\epsilon}$ which results in an extra factor $x^{-2\epsilon}$.
In this case the master integral is 
\begin{equation}
\int  _{0}^{1} {\rm d} x J(x) x^{n-2\epsilon}  =  B(n+2-4\epsilon,2-2\epsilon).
\end{equation}

\section {Applications}
\label{sec:applic}
In this section we  discuss some 
applications of the techniques described above.

First, we  give a complete result for the 
${\cal O}(\alpha _ s^2)$ correction to the 
differential semileptonic decay width of the $b$--quark
at the maximal recoil point.  We  next re-derive the 
BLM
${\cal O}(\alpha _s ^2)$ corrections to the quark top decay width
into massless $W$--boson and $b$ quark 
with the help of the expansions
presented above. As will be explained below, the expansion parameter
in this case is unity. 
It not obvious that the techniques described
above can be of any use there. 
Therefore it is interesting to
check how the procedure works in such extreme limit.  Finally, we 
comment on the applications to zero recoil sum rules.

\subsection{Two--loop QCD correction to semileptonic
$b$--decay at maximal recoil} 
\label {qmax}

The measurement of the inclusive semileptonic decay width 
permits a  determination of the  CKM matrix
parameter $|V_{cb}|$ with a small theoretical uncertainty
\cite {Shifman94,Neubert97}.
The  magnitude of the perturbative
corrections ${\cal O}(\alpha _s ^2)$ to this quantity
has been subject of discussions in the recent literature.
We addressed this problem  in ref.~\cite{MaxRec} where the exact
calculation of the ${\cal O}(\alpha _s ^2)$ corrections 
at maximal recoil
was used to estimate the total 4th order correction to the $b$ quark
semileptonic width.
Here we present more complete results
of that calculation.

We
consider  semileptonic decay of the $b$--quark
$b \to c l \nu _l$. The momentum carried away
by leptons is denoted by $q$.
We write differential semileptonic decay
width of the decay $b \to c l\nu _l$ at $q^2=0$ as
\begin{equation}
\left[ \frac {{\rm d}\Gamma _{\rm sl}}{{\rm d}q^2} \right]_{q^2=0}\;=
\;\Gamma _0 \left[\Delta _{\rm Born} + 
\frac {\alpha _s}{\pi}C_F \Delta _1+ \left(\frac {\alpha _s}{\pi}\right)^2
C_F \Delta _2 \right]
\label{eq:param}
\end{equation}
where $\Gamma _0 = \frac {G_F^2m_b^3}{96\pi^3}|V_{cb}|^2$.
$\Delta _{\rm Born} = (1-m_c^2/m_b^2)^3$ and $\Delta _1$ is also known
in a closed analytical form \cite{jk2,sa92b}. $\Delta _2$ is the new
result which gives ${\cal O}(\alpha _s ^2)$ correction to the
differential
semileptonic decay width of the $b$-quark at the point of 
zero invariant mass of leptons.

For the purpose of the presentation we write the 4th order correction as
\begin{equation}
\Delta _2 = \delta ^3 \left[(C_F-C_A/2)\Delta _F +C_A \Delta _A
+T_RN_L \Delta _L +T_R \Delta _H \right].
\label{eq:deltas}
\end{equation}

In the above equation $\delta = (1-m_c/m_b)$ is the expansion
parameter. $\Delta _H$ describes the contribution of the massive
$b$ and $c$ quark loops. 

For the $SU(3)$ group the color factors are $C_A = 3$, $C_F = 4/3$,
$T_R=1/2$. $N_L$ is the number of light quark flavors
whose masses were neglected.

In ref. \cite {MaxRec} we have presented analytical results
for the functions $\Delta _{F,A,L,H}$ up to $\delta ^4$, though
actual calculations have been done up to $\delta ^8$.
Our complete  formulas for these functions are presented in Appendix
\ref{app:8power}.

\subsection{BLM ${\cal O}(\beta _0 \alpha _s ^2)$
corrections to the decay width of the top quark}

As another application of the above techniques we consider the
decay of the top quark $t \to W b$. It is well known that this decay
width (at least at the Born and the one--loop level)
can be well approximated by neglecting the masses of the
$W$--boson and the $b$--quark.
In such case, since $W^2=0$,
the techniques presented in the main part of this paper can be readily
applied.  
In particular, the formula (\ref{eq:param}) can be rewritten in such
form that it gives the correction to the two-body decay width $t\to
QW$ (where $Q$ is a heavy quark): 
\begin{equation}
\Gamma(t\to QW) = \widetilde \Gamma_0 
\left[\Delta _{\rm Born} + 
\frac {\alpha _s}{\pi}C_F \Delta _1+ \left(\frac {\alpha _s}{\pi}\right)^2
C_F \Delta _2 \right].
\end{equation}
The $\Delta_i$ functions here are the same as in eq.~(\ref{eq:param}),
with $m_b$ and $m_c$ replaced by $m_t$ and $m_Q$.
The problem, however, is that the
procedures described above were based on the expansion of the
rate in the mass difference of the final and initial quarks.
In the case of the top quark decay with a $b$ quark in the final
state this means that the expansion parameter
$\delta = (m_t - m_b)/m_t$ is close to 1 for realistic
values of $m_t$ and $m_b$. 

\begin{figure}[h]
\hspace*{-5mm}
\begin{minipage}{16.cm}
\[
\mbox{
\hspace*{10mm}
\begin{tabular}{cc}
\hspace*{10mm}
\psfig{figure=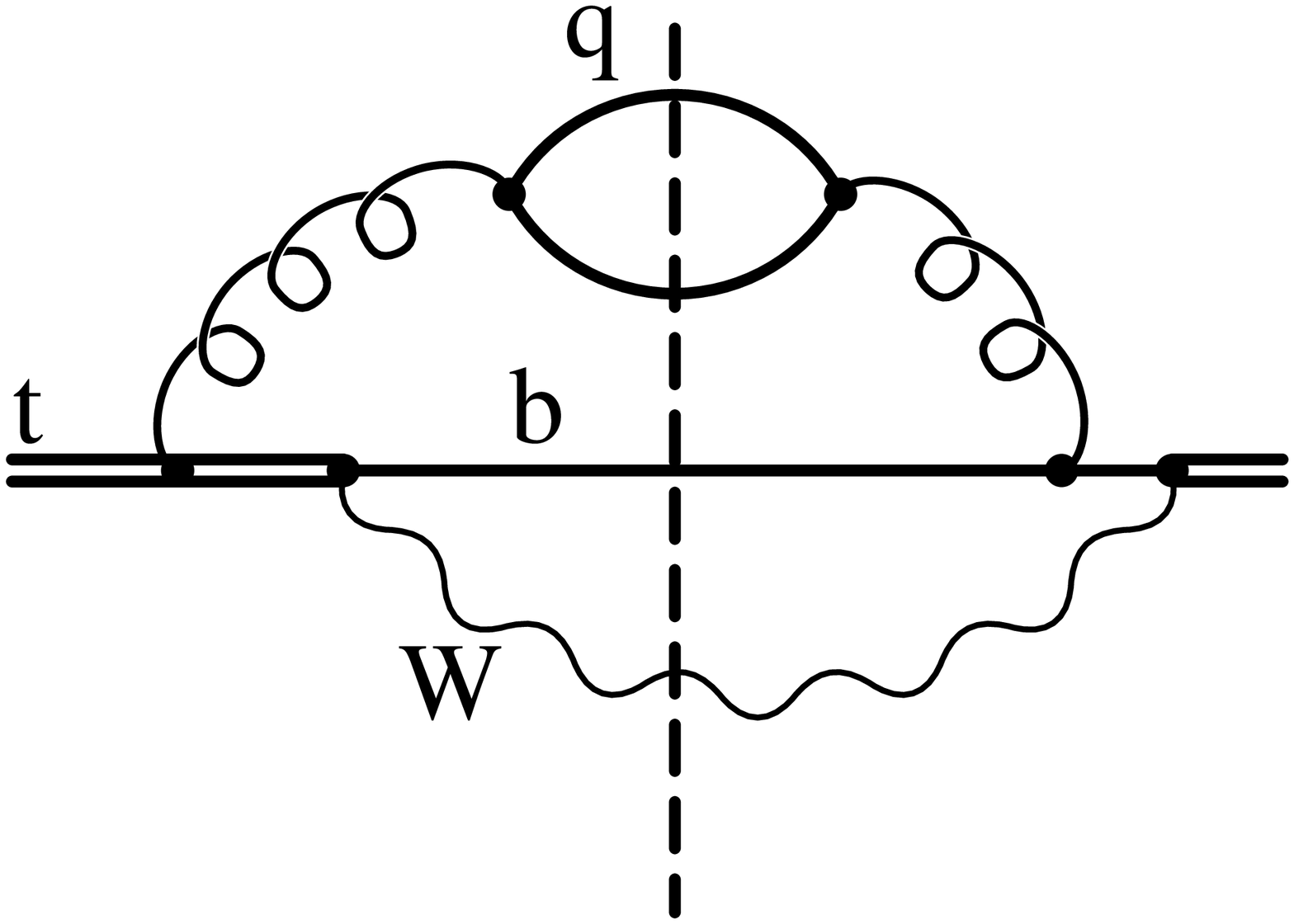,width=48mm,bbllx=210pt,bblly=410pt,%
bburx=630pt,bbury=550pt} 
&\hspace*{25mm}
\psfig{figure=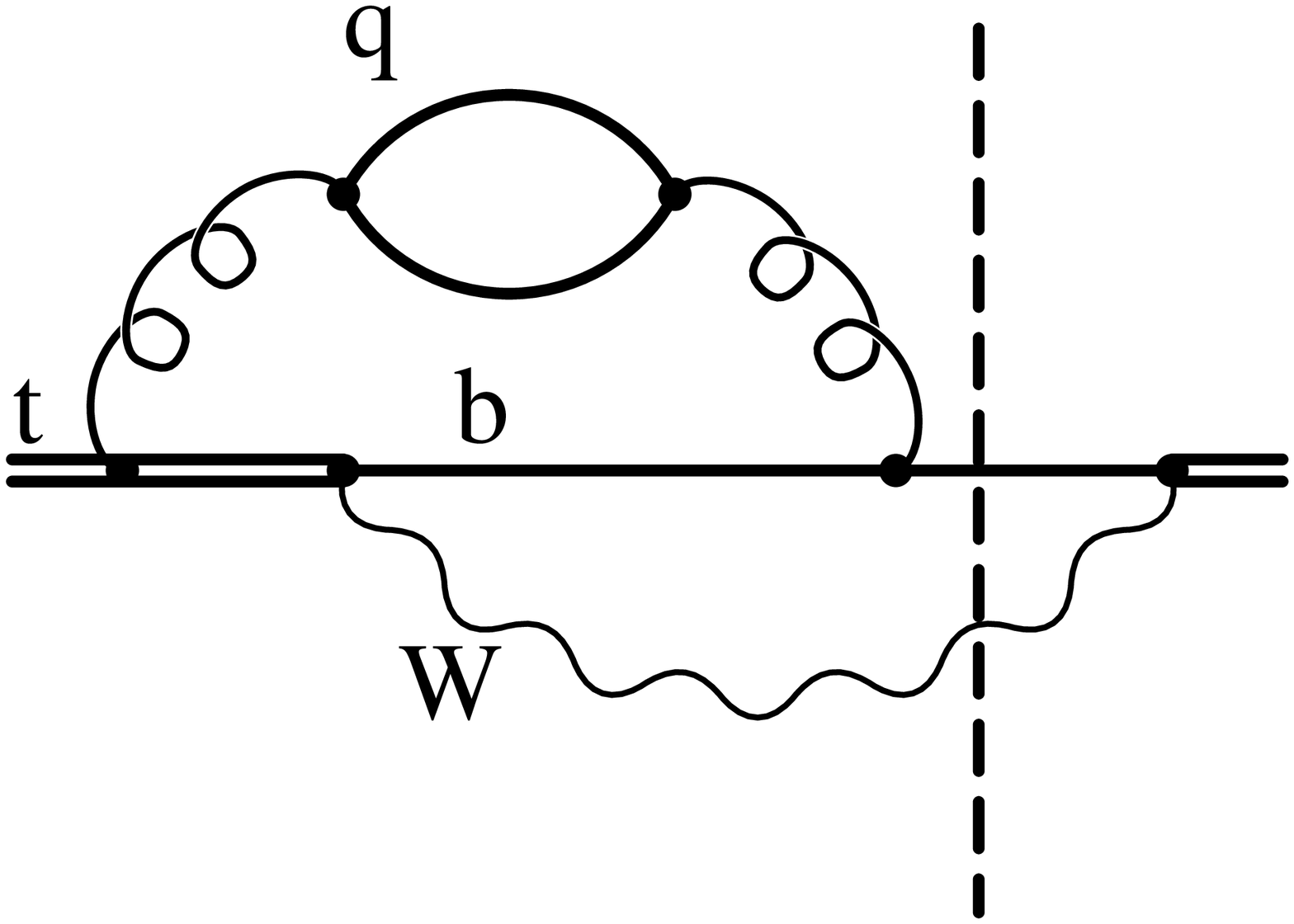,width=48mm,bbllx=210pt,bblly=410pt,%
bburx=630pt,bbury=550pt}
\\[25mm]
\hspace*{-12mm}
(a) & (b)
\end{tabular}}
\]
\end{minipage}
\caption{Examples of the light-quark loop corrections to the top quark
  decay.} 
\label{fig:blm}
\end{figure}

Little is known about convergence properties of the series described
in this paper.  Fortunately, part of the  ${\cal O}(\alpha _s ^2)$
correction to the top quark width is known exactly; it is the
contribution of the massless quarks, calculated
for $m_W = m_b = 0$ in \cite {smith,ac95a} (the relevant diagrams are
shown if Fig.~\ref{fig:blm}). 
We can use that limiting case to check if we can reproduce it with our
techniques.

The exact formula for the massless quark correction reads (in the
$\overline {\rm MS}$--scheme and 
for one generation of massless fermions)
\begin{eqnarray}
\Gamma ^{{\rm ferm}}(t\to Wb) &=& \left( \frac {\alpha _s}{\pi} \right)^2 
\frac {C_FT_R}{16} \widetilde\Gamma _0 \left [
2\zeta (3) +\frac {23}{9} \zeta (2) -\frac {8}{9}\right ],
\nonumber \\
\widetilde\Gamma _0 &=& \frac {G_F~m_t^3}{8\; \sqrt {2}\;\pi}.
\end{eqnarray}

It should be mentioned that  the diagrams with real or virtual
massless fermions represent the simplest case for the
described algorithms.  The reason is their simple planar topology,
which allows the computer programs to work very fast. 
For this particular type of diagrams
we can expand the width up to a very high power in $\delta$.

We write the result of this expansion as
$$
\Gamma ^{{\rm ferm}} (t \to b W ) =\left( \frac {\alpha _s}{\pi} \right)^2 
C_F~T_R \widetilde\Gamma _0 ~ F(\delta),
$$
where the function $F(\delta)$ is given as a series of powers and logs
of $\delta$; we have  computed this expansion up to terms 
of the order $\delta^{21} $.  The first few terms
of that expansion are given by the function $\Delta _L$ in the appendix.

The  numerical value of the exact result is
$$
F(1)=\frac {1}{16} \left (2\zeta (3) +\frac {23}{9} \zeta (2) -\frac {8}{9}
\right)=
0.3574\ldots.
$$
The values of the approximate result 
$F(\delta)$  for $\delta =1$ are, for three numbers $N$ of the summed
terms, 
\begin{eqnarray}
N  &=& 21,\qquad  F(1) = 0.3266,
\nonumber \\
N  &=& 15,\qquad  F(1) = 0.3210,
\nonumber \\
N  &=& 11,\qquad F(1) = 0.3176.
\end{eqnarray}

Comparing these numbers with the exact result 
we see that already the first 11 terms
of the expansion
give the accuracy of about $10\;\%$.
Unfortunately, the accuracy does not improve significantly with the
growing  number of terms in the expansion.  This slow convergence of
the series is caused by the  logarithms of the small ($b$ quark)
mass.  Although these terms are suppressed by powers of that mass (and
should give no contribution to the final result if $m_b=0$),
they spoil the convergence of the expansion.  It would be very useful
(and we think it is possible)
to find a systematic way of eliminating those parts of the integrands
which give rise to these logs.

On the other hand, from the perspective of practical applications
(like top physics at the Next Linear Collider) it would be sufficient
to know the two--loop QCD correction to the top quark decay width even
with $10$ percent accuracy.  It is tempting to apply our algorithms to
this problem and extrapolate the result to the point $\delta
= 1$; on the other hand, at present we do not know any reliable
method for estimating the uncertainty of such result.

\subsection{Zero recoil sum rules for $b \to c$ transition
with ${\cal O}(\alpha _s ^2)$ accuracy}

Another useful application of the techniques described above is connected
with the so-called zero recoil (ZR) sum rules \cite{Shifman94}.
In this subsection we would like to discuss  this point. A more
detailed
discussion of ZR sum rules with the 
${\cal O}(\alpha _s ^2)$ corrections will be given in a future
publication \cite{CzarMelUral97}.

The ZR sum rules are  important  for  the estimates
of the zero recoil transition form factors such as 
$F_{B \to D^*}$ for $B \to D^*$ transition or $F_{B \to D}$ for 
$B \to D$ transition. In turn, the formfactor $F_{B \to D^*}$ is a
crucial ingredient for $|V_{cb}|$ determination from the exclusive
semileptonic $B \to D^* l \nu _l$ decays. 

The sum rules for exclusive heavy--to--heavy flavor transitions are
based on the operator product expansion (OPE) of the hadronic
amplitudes in terms of the inverse quark masses. The zeroth order term
in this expansion is the parton model where  free quarks are
substituted for real hadrons both in initial and final state. Here we
disregard the non-perturbative corrections and discuss how the
perturbative corrections ${\cal O}(\alpha _s ^2)$ to the sum rules can
be evaluated.

We consider a transition of a $b$ quark at rest to a $c$ quark and
massless partons which occurs under the influence of the external
current $J_{\mu}$.
The momentum carried away by the external current is $q = (q_0, 0)$, where
$$
q_0 = m_b - m_c - \epsilon,\qquad \epsilon \le \mu \ll m_b,m_c.
$$

The quantity which is of primary importance for the perturbative
corrections to ZR sum rules can be schematically written as:
$$
\int \limits _{0}^{\mu} {\rm d}\epsilon\; w_{J}(\epsilon),~~~
w_{J}(\epsilon) = \Gamma _{J} (b \to c+X\;
|\; q_0).
$$
In this equation we have shown explicitly the dependence of the transition
rate $\Gamma _{J} (b \to c\; |\;q_0)$ both on the external momenta transfer
$q_0$ and on the Lorentz structure of the current.

If $\epsilon =0$, the transition is elastic; the final state is a single
$c$--quark at rest. The ${\cal O}(\alpha _s ^2)$ corrections in this
case reduce to the renormalization of the external current $J_{\mu}$.
For vector and axial currents they were calculated in
\cite{zerorecoil,zerorecoilA}.

On the other hand, if $\epsilon \ne 0$ the $c$ quark starts moving and can
radiate.  The second order corrections ${\cal O}(\alpha _s)$ were
calculated in \cite {Shifman94,Korner:1996,Kapustin96}.  Aiming at the
fourth order, i.e. at ${\cal O}(\alpha _s^2)$ accuracy, 
one has to consider the
final state with the $c$--quark and two real gluons or light quarks
and also the ${\cal O}(\alpha _s)$ correction to the single gluon
emission in $b \to c$ transition.

Due to the hierarchy of scales, $\epsilon < \mu \ll m_b,m_c$, 
the final $c$--quark
is moving slowly by definition; this  is therefore a nice place
where the techniques described in this paper can be applied.
In particular, an algorithm for performing eikonal expansions 
with the subsequent integration over the phase space
appears to be very useful here. This topic will be discussed in detail
in ref.~\cite {CzarMelUral97}.

\section{Conclusion}

In this paper we have described techniques used in
the calculation of the ${\cal O}(\alpha _s ^2)$ corrections to the
semileptonic decay width $b \to c l \nu _l$ at maximal recoil
\cite {MaxRec}.
 
The technical tool we used for that calculation is an expansion of the
decay rate in powers and logarithms 
of the mass difference between the initial and
final quarks. We presented a detailed discussion of the algorithm,
which enables one to construct such expansion.  We treated virtual
corrections, emission of one gluon, and emission of two gluons
separately.  Therefore, these algorithms can be used for the analyses
of less inclusive quantities than the total decay rate, at least in
principle.

In case of two--loop virtual corrections and  emission of two real 
gluons, the expansion in $\delta = (m_b-m_c)/m_b$ is a Taylor
expansion. In the case of one--loop corrections to the amplitude
of single gluon emission in $b$-decay, the Taylor expansion is
insufficient. An appropriate method is provided by the eikonal
expansions, recently introduced in refs. \cite {Smirnov96,CzarSmir96}.
When this procedure is used, a new 
type of Feynman integrals appears.
These integrals and methods which were used for their evaluation
were described here in some detail.

The whole construction works well if the mass difference between
initial and final state quarks is not too large.  This is
the case for the semileptonic $b \to c$ transitions, where the
expansion parameter $(m_b-m_c)/m_b \sim 0.7$.
In this case we calculated the expansion up to the eleventh power of
$\delta$; the estimated accuracy of the final result is better than
$1 \%$.

There is, however, a
number of other applications where the initial quark is significantly 
heavier
than the final one.  It is not obvious to what extend the present
method can be useful in such situation.  
There is an indication, however, that our procedures give
meaningful results even in that limit.
As an example, we analyzed the 
light quark 
corrections  to the width of the top quark decay  into massless $W$ boson
and a $b$-quark. We have shown that the first several terms of the
expansion in
$\delta _t = (m_t-m_b)/m_t$ approximate the known exact result 
with a $10 \%$ accuracy.

Finally, we have argued  that the same techniques can be applied to
the corrections to the zero recoil sum rules for the heavy flavor
transitions. 

\section*{Acknowledgments}
We are grateful to K.~G.~Chetyrkin  and N.~G.~Uraltsev 
for many helpful discussions.
We would like to thank Prof. J.~H.~K\"uhn for his interest in this
work and support. 
This research has been supported by BMBF 057KA92P  and by
Graduiertenkolleg ``Teilchenphysik'' at the University of Karlsruhe.


\begin{thebibliography}{10}


\bibitem{RPP}
{Particle Data Group}, Phys. Rev. {\bf D54},  1  (1996).

\bibitem{Shifman94}
M. Shifman, N.~G. Uraltsev, and A. Vainshtein, Phys. Rev. {\bf D51},  2217
  (1995), erratum: ibid. D52, 3149 (1995).

\bibitem{Neubert97}
M. Neubert, preprint hep-ph/9702375, 
 to appear in  A. Buras and M. Lindner (eds.),
 {\em Heavy Flavours}, 2nd edition, (World Scientific, Singapore). 


\bibitem{zerorecoil}
A. Czarnecki, Phys. Rev. Lett. {\bf 76},  4124  (1996).

\bibitem{zerorecoilA}
A. Czarnecki and K. Melnikov, hep-ph/9703277, submitted to Nucl. Phys. B
  (unpublished).

\bibitem{MaxRec}
A. Czarnecki and K. Melnikov, Phys. Rev. Lett. {\bf 78},  3630  (1997).

\bibitem{topfermion}
A. Czarnecki,  in {\em New computing techniques in physics research {IV}},
  edited by B. Denby and D. {Perret-Gallix} (World Scientific, Singapore,
  1995), pp.\ 319--323, Proceedings of the International Workshop 
  AINHEP 1995, Pisa, Italy.

\bibitem{Smirnov96}
V.~A. Smirnov, Phys. Lett. {\bf B394},  205  (1997).

\bibitem{CzarSmir96}
A. Czarnecki and V.~A. Smirnov, Phys. Lett. {\bf B394},  211  (1997).

\bibitem{che81}
K.~G. Chetyrkin and F. Tkachov, Nucl. Phys. {\bf B192},  159  (1981).

\bibitem{jk2}
M. Je{\.z}abek and J.~H. K{\"u}hn, Nucl. Phys. {\bf B314},  1  (1989).

\bibitem{sa92b}
A. Czarnecki and S. Davidson, Phys. Rev. D {\bf 48},  4183  (1993).

\bibitem{smith}
B.~H. Smith and M.~B. Voloshin, Phys. Lett. {\bf B340},  176  (1994).

\bibitem{ac95a}
A. Czarnecki, Acta Phys. Pol. {\bf B26},  845  (1995).

\bibitem{Korner:1996}
J.~G. K\"orner, K. Melnikov, and O. Yakovlev, Z. Phys. {\bf C69},  437  (1996).

\bibitem{Kapustin96}
A. Kapustin, Z. Ligeti, M.~B. Wise, and B. Grinstein, Phys. Lett. {\bf B375},
  327  (1996).

\bibitem{CzarMelUral97} A.~Czarnecki, K.~Melnikov,~N.~G.~Uraltsev, 
in preparation.

\end{thebibliography}

\appendix
\section{Results for the maximal recoil}
\label{app:8power}

In this appendix we present the first nine terms of the expansion of
the coefficient functions for the $b$ quark decay rate at maximal
recoil, as defined in (\ref{eq:deltas}):
\begin{eqnarray}
\lefteqn{\Delta_{A}  = 
       - {355 \over 36} + {2 \over 3} \pi^2
       + \delta   \left( {89 \over 8} - \pi^2 \right)
}
\nonumber \\ &&
       + \delta^2   \left(  - {2422517 \over 32400} + {1708 \over 45}
       \ln(2\delta) - {44 \over 9}  
         \ln^2(2\delta) + {8 \over 9} c_1 + {257 \over 90} \pi^2 \right)
\nonumber\\ &&
       + \delta^3   \left( {2956607 \over 64800} - {854 \over 45}
       \ln(2\delta) + {22 \over 9}  
         \ln^2(2\delta) - {4 \over 9} c_1 - {307 \over 180} \pi^2 \right)
\nonumber\\ &&
       + \delta^4   \left(  - {5789957 \over 1323000} + {4663 \over
         4725} \ln(2\delta) + {2 \over 5}  
         \ln^2(2\delta) + {4 \over 45} c_1 + {412 \over 1575} \pi^2
       \right) 
\nonumber\\ &&
       + \delta^5   \left(  - {4413299 \over 1984500} + {13063 \over
         9450} \ln(2\delta) + {1 \over 5}  
         \ln^2(2\delta) + {233 \over 4725} \pi^2 + {2 \over 45} c_1
       \right) 
\nonumber\\ &&
       + \delta^6   \left(  - {303218053 \over 222264000} + {290963
         \over 264600} \ln(2\delta) 
          + {251 \over 1260} \ln^2(2\delta) + {3349 \over 132300}
          \pi^2 + {23 \over 630} c_1 \right) 
\nonumber\\ &&
       + \delta^7   \left(  - {510931331 \over 444528000} + {104561
         \over 105840} \ln(2\delta) 
          + {167 \over 840} \ln^2(2\delta) + {4573 \over 264600} \pi^2
          + {41 \over 1260} c_1 \right) 
\nonumber\\ &&
       + \delta^8   \left(  - {179575376107 \over 172889640000} +
       {48009667 \over 52390800} 
          \ln(2\delta) + {1859 \over 9450} \ln^2(2\delta) + {18679
            \over 1496880} \pi^2 + {4 \over 135} c_1 \right) 
\nonumber\\
\lefteqn{\Delta_{F}  =
       - {23 \over 6} + {8 \over 3} c_2 + {8 \over 3} \pi^2
       + \delta   \left( {23 \over 4} - 4 c_2 - 4 \pi^2 \right)
       + \delta^2   \left( {1697 \over 360} - {8 \over 3} \ln(2\delta)
       + {22 \over 5} c_2 + {359 \over 135} \pi^2 
          \right)
} 
\nonumber\\ &&
       + \delta^3   \left(  - {3347 \over 720} + {4 \over 3}
       \ln(2\delta) - {23 \over 15} c_2 - {179 \over 270}  
         \pi^2 \right)
\nonumber\\ &&
       + \delta^4   \left( {4957991 \over 396900} - {1460 \over 189}
         \ln(2\delta) + {16 \over 9}  
         \ln^2(2\delta) - {139 \over 600} \pi^2 + {2 \over 7} c_2 \right)
\nonumber\\ &&
       + \delta^5   \left( {1803791 \over 793800} - {394 \over 189}
         \ln(2\delta) + {8 \over 9}  
         \ln^2(2\delta) - {139 \over 1200} \pi^2 + {1 \over 7} c_2 \right)
\nonumber\\ &&
       + \delta^6   \left( {352922723 \over 142884000} - {55093 \over
         28350} \ln(2\delta) + {4 \over 5} 
          \ln^2(2\delta) - {35671 \over 352800} \pi^2 + {29 \over 252}
         c_2 \right) 
\nonumber\\ &&
       + \delta^7   \left( {221825423 \over 95256000} - {34273 \over
         18900} \ln(2\delta) +  
      {34 \over 45} \ln^2(2\delta) - {22049 \over 235200} \pi^2 + {17
         \over 168} c_2 \right) 
\nonumber\\ &&
       + \delta^8   \left( {93748812373 \over 43222410000} - {3067283
         \over 1819125}  
         \ln(2\delta) + {1133 \over 1575} \ln^2(2\delta) - {10703293
         \over 119750400} \pi^2 +  
         {211 \over 2310} c_2 \right)
\nonumber\\
\lefteqn{\Delta_{L}  =       
        {14 \over 9}
       - \delta 
       + \delta^2   \left( {82217 \over 4050} - {544 \over 45}
         \ln(2\delta) + {16 \over 9} \ln^2(2\delta) 
          - {16 \over 27} \pi^2 \right)
} 
\nonumber\\ &&
       + \delta^3   \left(  - {103667 \over 8100} + {272 \over 45}
         \ln(2\delta) - {8 \over 9}  
         \ln^2(2\delta) + {8 \over 27} \pi^2 \right)
\nonumber\\ &&
       + \delta^4   \left( {1322183 \over 496125} - {2404 \over 1575}
         \ln(2\delta) + {8 \over 45}  
         \ln^2(2\delta) - {8 \over 135} \pi^2 \right)
\nonumber\\ &&
       + \delta^5   \left( {1653341 \over 1984500} - {1202 \over 1575}
         \ln(2\delta) + {4 \over 45}  
         \ln^2(2\delta) - {4 \over 135} \pi^2 \right)
\nonumber\\ &&
       + \delta^6   \left( {18480089 \over 27783000} - {4831 \over
         7350} \ln(2\delta) + {23 \over 315}  
         \ln^2(2\delta) - {23 \over 945} \pi^2 \right)
\nonumber\\ &&
       + \delta^7   \left( {33623843 \over 55566000} - {26651 \over
         44100} \ln(2\delta) + {41\over 
         630} \ln^2(2\delta) - {41 \over 1890} \pi^2 \right)
\nonumber\\ &&
       + \delta^8   \left( {1620326143 \over 2881494000} - {1053959
         \over 1871100} \ln(2\delta) 
          + {8 \over 135} \ln^2(2\delta) - {8 \over 405} \pi^2 \right)
\nonumber\\
\lefteqn{\Delta_{H}  =
        {460 \over 9} - {16 \over 3} \pi^2
       + \delta   \left(  - 74 + 8 \pi^2 \right)
       + \delta^2   \left( {9821 \over 81} - {344 \over 27} \pi^2 \right)
}
\nonumber\\ &&
       + \delta^3   \left(  - {33883 \over 810} - {32 \over 9}
         \ln(2\delta) + {136 \over 27} \pi^2 \right) 
\nonumber\\ &&
       + \delta^4   \left( {3754 \over 405} - {154 \over 135} \pi^2 \right)
       + \delta^5   \left( {174907 \over 28350} - {88 \over 135}
         \ln(2\delta) - {77 \over 135} \pi^2 \right) 
\nonumber\\ &&
       + \delta^6   \left(  - {853453 \over 18900} - {88 \over 135}
         \ln(2\delta) + {17489 \over 3780}  
         \pi^2 \right)
       + \delta^7   \left(  - {56241287 \over 793800} - {586 \over
         945} \ln(2\delta) + {54623\over 
         7560} \pi^2 \right)
\nonumber\\ &&
       + \delta^8   \left(  - {35304151 \over 297675} - {556 \over
         945} \ln(2\delta) 
+ {39073\over 3240} \pi^2 \right)
\nonumber \\
\end{eqnarray}
with $c_1={21\over 2}\zeta_3-\pi^2\ln\left(2\delta\right)$
and $c_2={3\over 2}\zeta_3-\pi^2\ln 2$.

\end{document}